# Investigation of tantalum films growth for coplanar resonators with internal quality factors above ten million

E.V. Zikiy,[1,2] N.S. Smirnov,[1,2] E.A. Krivko,[1] A.R. Matanin,[1] A.I. Ivanov,[1] E.I. Malevannaya,[1] V.I. Polozov,[2] S.V. Bukatin,[1] D.A. Baklykov,[1] I.A. Stepanov,[1] S.A. Kotenkov,[1] S.P. Bychkov,[1] D.A. Fokin,[1] I.A. Ryzhikov,[1] A.V. Andriyash[2] and I.A. Rodionov[1,2,*]

[1]Shukhov Labs, Quantum Park, Bauman Moscow State Technical University, Moscow, 105005, Russia
[2]Dukhov Automatics Research Institute (VNIIA), Moscow 127055, Russia
*email: irodionov@bmstu.ru

## ABSTRACT

Alpha-tantalum on silicon is a promising platform for high-coherence superconducting quantum circuits. However, the growth mechanism of alpha-tantalum on silicon remains poorly understood. We present a comprehensive study on alpha-tantalum films growth on various substrate. The decisive role of a substrate material Debye temperature on phase selection mechanism in tantalum films growth is experimentally confirmed, contradicting the prior assumptions on substrate temperature influence. Crucially, we confirm that alpha-tantalum starts growing only after a 7-10 nm thick beta-tantalum sublayer. It results in ranging the critical temperature of α-Ta films from 3.77 K to 4.39 K for the total thickness from 20 to 150 nm, respectively. Finally, we compared high-quality Al and Ta coplanar resonators on silicon, demonstrating compact tantalum resonators (4/10.5/4 μm) with an internal quality factor exceeding 10 million at single-photon excitation powers.

## I. INTRODUCTION

Scalable quantum computers both noisy intermediate-scale and error-corrected require high fidelity of two-qubit gates well above 99.9% threshold [1], [2], [3]. Limited coherence of superconducting qubits is one of the key sources of gate errors [4]. A major component of qubit decoherence – relaxation – is currently, in the state of the art of superconducting qubits, primarily caused by dielectric losses [5], [6]. There are several ways to reduce dielectric losses: optimizing qubit design, improving fabrication processes, and new material platforms research [7], [8], [9]. Tantalum in its alpha phase has recently emerged as a potential candidate foe next generation superconducting circuits platform due to its native low loss and high resistance to aggressive TLS removing techniques. Transmon qubits with relaxation times exceeding 300 μs and 500 μs have been demonstrated, achieving record quality factors [10], [11]. In these papers, sapphire substrate was used due to low dielectric loss. However, sapphire is CMOS-incompatible and cannot be used in 300-mm wafers fabrication and 3D integration involving through-silicon vias (TSVs) [12], [13]. Silicon substrates are essential in superconducting quantum processors technology with 1000 qubits and above. Yet, the mechanism of growing thin alpha-phase tantalum on silicon remains poorly understood.

Tantalum thin films support two phases: BCC α-Ta and tetragonal β-Ta [14]. While β-Ta is used in resistors [15] and heaters [16] due to pretty high resistivity, α-Ta serves as a diffusion barrier in Cu/Si interconnects [17], [18], [19], protective coatings [20], [21], capacitor plates [22], [23], wear-resistant layers [24], [25], and anti-reflection coatings [26], [27]. For superconducting quantum computing applications, only α-Ta is suitable due to low critical temperature ($T_C$) of β-Ta's [28], [29]. α-Ta typically forms when substrates are heated above 400°C [30], [31], [32]. β-Ta was shown to grow on room temperature amorphous substrates [30], oxidized surfaces [33], [34] or oxygen-contaminated interfaces [35] suggesting that heating promotes α-phase formation by desorbing adsorbates. However, α-Ta growth without heating on silicon substrates remains unproven, even with chemical or plasma cleaning. This implies a deeper thermal influence on phase selection, requiring further study of α-Ta nucleation mechanisms.

Here, we investigate a phase selection mechanism in tantalum thin films and confirm its strong correlation with a substrate Debye temperature. Through a detailed research of the initial stages of tantalum film growth on silicon, we experimentally demonstrate that α-tantalum growth initiates only after a 10 nm thick β-phase tantalum sublayer formation. We demonstrate the correlation between the initial growth stage and the grain structure of thicker α-phase films and propose magnetron sputtering process resulting an ultralow roughness. We also stated that pre-cleaning of substrate adsorbates is of a secondary order factor governing the growth of α-tantalum over β-tantalum. To ensure our α-Ta films quality we report superconducting coplanar waveguide (CPW) tantalum resonators on silicon with an internal quality factor exceeding 10 million in the single-photon regime. CPW resonators remains a reliable method for assessing dielectric losses in superconducting quantum circuits, providing valuable insights into their origin. To date, internal quality factors (at single-photon excitation powers of $4.4\times10^6$ [36], $2.1\times10^6$ [37] and $1.1\times10^6$ [38] have been demonstrated for α-tantalum CPW resonators on silicon substrates. Notably, the intrinsic quality factor of tantalum resonators on sapphire substrates does not surpass these values [39], [40], [41], [42]. Implementing 2 um deep silicon trenching, we achieved high-quality factor tantalum CPW resonators with a compact cross-section of 4/10.5/4 μm. Through systematic interface engineering via trenching and post-processing – with direct comparison to aluminum reference resonators – we prove that the critical substrate-metal interface remains unaffected by 10 nm thick β-Ta sublayer. Consequently, the sublayer does not limit the coherence of superconducting quantum circuits. This detailed study of tantalum film growth and their structural properties could be of a great importance for tantalum-based superconducting quantum circuits platform.

## II. RESULTS AND DISCUSSION

### A. α-Ta films structure

Superconducting tantalum thin films are usually magnetron sputtered in ultrahigh vacuum at substrate temperatures up to 800°C. In this work we use high-resistivity silicon substrates (25×25 mm, $\rho$ > 10 kΩ×cm) in all the experiments, which were cleaned in Piranha solution and dipped in HF to remove native oxide and passivate the surface (see Appendix A for details). All the samples are sputtered follow a unified deposition rate of 0.6 nm/s. We have improved tantalum film surface roughness by simultaneously optimizing the substrate temperature in the range from 400 to 800 °C and working pressure from 0.3 to 9.0 mTorr (see Appendix B for the details). The sputtering system is described in [43], its simplified schema is shown in Figure 4d (Appendix A).

In this study, we employed four-point probe resistivity mapping at 49 points across each substrate to differ α-phase and β-phase of tantalum films. Sometimes X-ray diffraction (XRD) can be unreliable for tantalum phase identification due to close peaks in the 2-theta diagrams, such as for α-Ta (110) ≈38.5° and β-Ta (202) ≈38.2° [33], which can lead to ambiguous interpretations [44], [45] when peaks are shifted due to internal stresses in the films. However, the most common orientation β-Ta (002) ≈33.5° is identified unambiguously. Electrical resistivity offers more robust alternative, as shown in prior studies [30], [36], [38], [46], [41]: pure α-phase exhibits $\rho_{Ta} \leq 25$ μΩ×cm, while pure β-phase has from 140 μΩ×cm to 170 μΩ×cm [47]. Surface morphology is characterized with dark-field optical microscopy, scanning electron microscopy (SEM), and atomic force microscopy (AFM). Crystalline structure is analyzed by means of SEM electron backscattered diffraction (EBSD), which has proven a reliable technique [48], [49]. Film surface roughness is measured with both stylus profiler with ultrasharp stylus (Rq) and AFM (Sq).

We observed that α-Ta is growing on silicon forming a distinctive flower-like grains (Figure 1), hereafter referred to as “flowers”. One can conclude an average grain size of 30-40 μm from a 600 × 800 $\mu m^2$ grain map and corresponding distribution histogram (Figure 1, b). EBSD mapping (Figure 1, d) confirms features as the individual grains. The grains exhibit uniform 110 orientation along the z-axis with random orientations in the x-y plane. Note, that the backscattered electron detector is places at an angle to the sample, which results in the apparent grain elongation. The light grain boundaries and centers of each "flower" can be defined with optical microscopy (Figure 1, g). Corresponding SEM images at similar magnification (Figure 1, h) show its detailed morphology. One can notice nanoridges at tantalum surface: the "petals" of the "flowers" (Figure 1, i), while the "background" is extremely smooth (Figure 1, j). Smooth areas have surface roughness of Sq ≈ 4.2 Å and nanoridge of Sq ≈ 16.9 Å (Figure 1, e). Nanocrystalline structure β-Ta films have much smoother surface (Figure 1, f) compared to α-Ta.

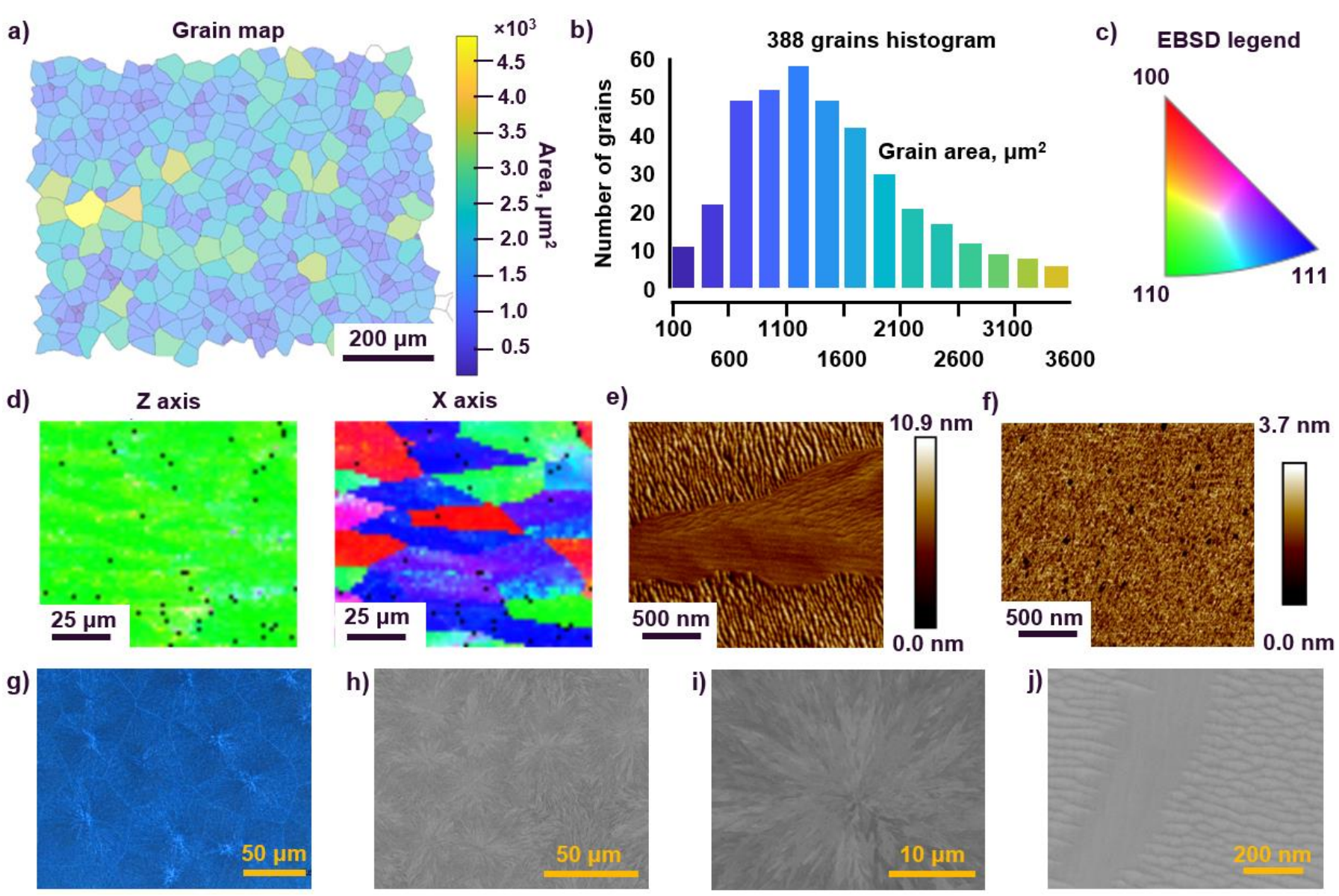

Figure 1. Characterization of a 150 nm thick α -Ta films on a silicon substrate: (a) Grain map (false color optical microscopy) and its distribution histogram (b). (c) Legend for EBSD maps. (d) SEM EBSD surface maps of 100×100 μm. (e) AFM image of α-Ta/Si and (f) β-Ta/Si at high magnification. (g) Dark-field optical microscopy of α-Ta film. (h) SEM

images of α-Ta grains at 500 X, the “flower” center at 3 kX (j), and smooth/nanoridge area at 100 kX (j).

We expect, that deposition of these exceptionally large grains likely originates from the initial film growth stage. However, the coexistence of smooth and nanoridge regions remains unexplained, with the region boundaries ranging from gradual to abrupt. Similar nanoridge structures have been reported in tungsten films [50] and α-Ta [51], where different oxidation mechanisms of these regions was observed. This structural heterogeneity and the associated variations in surface chemical activity may pose challenges for α-Ta/Si applications in nanoelectronics, particularly for devices requiring precise dimensional control [52].

### B. α-Ta films origin

The fact that tantalum grains have a single orientation along the z-axis, but are disoriented in the plane and have a radial, centered structure allowed us to assume that the film formation originates from some growth centers, which later become the "flowers" centers. To test this hypothesis several tantalum films were sputtered on silicon substrates at 500 °C with 5, 10, 15, 20, and 50 nm thicknesses. The resistivity dependence on the film thickness is shown in Figure 2, a: the 5-nm-thick and 10-nm-thick films $\rho_{Ta}$ obviously agrees with the tantalum β-phase. But with further increase in film thickness, $\rho_{Ta}$ becomes nearer to α-Ta typical value. We confirmed experimentally that the resistance of very thin β-Ta films does not differ significantly from the bulk resistance, which helps to avoid errors in calculating the resistance of the α-Ta layer on the β-Ta sublayer. The highest resistance was found for the 5 nm thick β-Ta film and was ≈ 160 μΩ×cm (see Appendix F for details).

Centric-like “flowers” grains growth can be visualized using dark-field optical microscopy of Ta films with thicknesses of 5, 10, and 15 nm, as shown in Figure 2, b. One can notice almost no structures on the surface of a 5-nm-thick films, the same as for 150-nm-thick β-Ta films. But single amount of the point-like growth centers starts appearing on the surface (which could look like defect at a first glance for 5-nm-thick films). For a 10-nm-thick film the number of growth centers become much larger and some of them start growing in the XY plane forming small “flowers”, which at 15 nm thickness occupy half of the entire film surface. SEM image of 15-nm-thick film (Figure 2, d) confirms dark-field optical microscopy results. Figure 2, e shows SEM image of the center of the growing spot, which illustrate the future "flower" is observed.

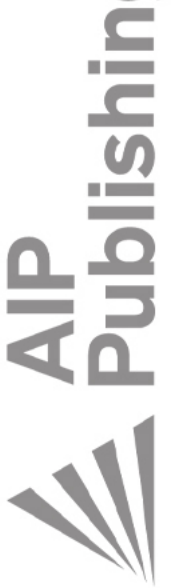

While for 20- and 50-nm-thick films the different regions are no longer detected. It should be noted, that there is no height step between smooth and growing "flower" surfaces according to angled high resolutions SEM of the boundary (Figure 2, f). The absence of island growth is also confirmed by the AFM data presented in Figure 9, c (see Appendix F for details). The height difference between the two regions according to the AFM data is only 0.87 Å and is due to the higher density of the α-phase, i.e., some shrinkage is present. We assume that the structure of the future "flower" grows upward in a cone, as schematically shown in Figure 2, a. One can also clearly see round crystalline α-Ta regions extracted from 15-nm-thick film EBSD-data (Figure 2, c); zero solutions (shown in black) correspond to the tantalum β-phase. β-Ta is not detectable as a crystalline material at all by EBSD, as even a 150 nm thick β-Ta film is also colored black in the EBSD-map, and, most importantly, the Kikuchi lines are completely absent. Note that a recent study [95] presented orientation maps of the cross-sections of the films with β-Ta defects, obtained using the scanning transmission electron microscopy (STEM) based automatic crystal orientation mapping method. The β-Ta regions in this study were also marked in black, without indicating their orientation. Our EBSD-results perfectly confirm the films resistivity measurements ($\rho_{Ta}$). It can be concluded, that when sputtering Ta at high temperatures on silicon first the sublayer is formed in β-phase with the thickness of 5-10 nm. Then the point-like growth centers of α-phase are formed on the surface of β-phase, which further expending in α-phase regions. The average distance between the α-phase regions centers is around 30-40 μm, resulting in a thicker film comprising large "flowers" grains.

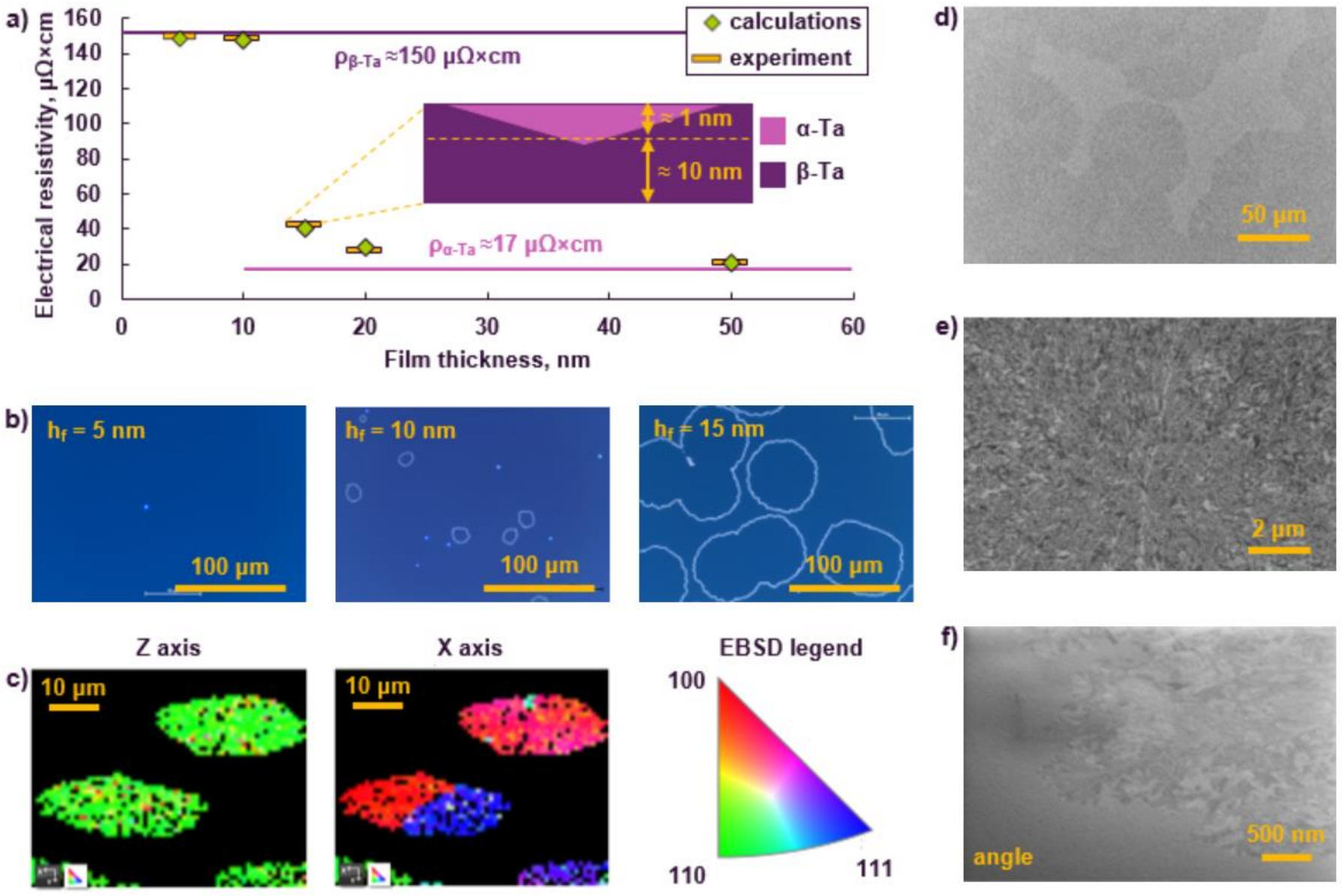

Figure 2. Characterization of the initial growth stage of an α-Ta film on a heated to 500 °C silicon substrate (a) Experimental and calculated resistivity of tantalum films as a function of thickness; the inset shows the schematic of the initial growth stage of α-Ta/Si films. (b) Dark-field optical images of Ta films on Si with thicknesses of 5, 10, 15 nm. (c) EBSD maps of 15 nm Ta/Si films and their legend (α-Ta is colored; β-Ta is black). (d) SEM-image of a 15 nm Ta/Si film. (e) SEM-image of the α-Ta region at the center of the 15 nm Ta/Si film. (f) SEM image taken at an angle showing of the boundary between the α-Ta (up right) and β-Ta (down left) regions in a 15 nm Ta/Si film.

The growth of an α-Ta film on an ultrathin β-Ta sublayer for Ta thin films sputtering on heated silicon substrates represents a unique growth mechanism, contrasting to conventional models, when the initial nucleation phase determines the entire film's structure. While the most studies suggest tantalum film growth in either pure α- phase or β-phase depending on sputtering conditions [30], [33], [34], [35], our findings align with in situ observations [31] showing β→α recrystallization during growth on heated glass. However, we have experimentally confirmed, that even for 20- and 50-nm-thick films with the whole surface detected as α-phase, their resistivity remains higher than 150-nm-thick α-Ta films (close to bulk properties), suggesting the initial 10 nm β-phase sublayer persists and does not recrystallize. We modeled the resistivity behavior assuming α- or β-phase as parallel conductors:

$$\rho_{Ta} = \frac{\frac{\rho_\beta}{10} \times \frac{\rho_\alpha}{(h-10)}}{\frac{\rho_\beta}{10} + \frac{\rho_\alpha}{(h-10)}} \times h, \quad (1)$$

where $\rho_\beta$ and $\rho_\alpha$ are the resistivity of 150 nm β-Ta and α-Ta films respectively, and $h$ is the film thickness in nanometers. The perfect agreement between calculated and experimental values (Figure 2, a) confirms that β-Ta remains as part of the thicker film. The sublayer is also visible in the focused ion beam (FIB) cross-sectional image shown in Figure 9, a (see Appendix F for details).

We note that the XRD-data for the mixed α+β Ta film presented in the study [95] indicate the presence of 10% β-Ta. In our case, the XRD-data for our films (see Appendix A for details) do not show of β-Ta. We assume that the β-Ta sublayer in our films is disordered, which is why it is not detected by XRD. This seems quite realistic, since β-Ta consists of topologically close-packed icosahedral structures, which are also found in amorphous Ta [54]. Since the β-Ta sublayer is likely stabilized by interdiffusion of Ta and silicon, the β-Ta structure is quite disordered. Furthermore, this layer could be either α-Ta, or β-Ta, or amorphous Ta. For α-Ta, the electrical resistance is too high, while for amorphous Ta, the resistance is too low [107]. Therefore, it is most likely β-Ta.

Despite the existence of a thin sublayer of β-Ta, when the substrate is heated, the alpha phase of Ta is formed. We attempted to uncover the mechanism of phase selection in thin tantalum films. Our substrate pretreatment experiments (see Appendix A for details) reveal that substrate heating provides more than just adsorbates removal - it creates essential energetic conditions for α-phase formation in silicon, sapphire and $SiO_2$ substrates. Both molecular dynamics simulations [53], [54] and experimental studies [55] show that Ta-melt requires anomalously low cooling rates of $10^{11}$-$10^{13}$ K/s to form amorphous metallic glass. The β-Ta phase possesses a topologically close-packed structure (TCP crystal [54], Frank Casper σ-phase [56]) composed of icosahedral clusters [53], structurally similar to Ta metallic glass [57]. This explains why β-Ta forms at high cooling rates, though lower than required for glass formation [54]. It is well known that deposited atoms lose excess energy in only a few periods of the substrate lattice vibration [58], [59], since the main mechanism of energy relaxation between adatoms and the surface is phonon conversion, although there are other secondary factors such as exchange into rotational modes or electronic excitations. In this case, the maximum frequency of lattice vibrations during adatom capture is determined by the Debye temperature ($\theta_D$) of the substrate material, and we assume that the captured adatom leads to the generation of phonons with the maximum frequency [60], [61]. Analysis of the relationship between $\theta_D$ and resulting Ta phase (Table 1) reveals that β-Ta preferentially forms on high-$\theta_D$ substrates at room temperature. This correlation suggests, that substrates with high $\theta_D$ enable rapid thermalization of deposited Ta atoms, favoring the β-phase formation. When the substrate is heated, the reduced thermal gradient between deposited adatoms and the substrate decreases the effective cooling rate, resulting in α-Ta growth. Notably, oxides typically have higher Debye temperature [62], [63], [64] explaining the distinct phase selection on different underlayers: α-Ta forms on clean as-deposited Nb, Mo, or Al surfaces, while their exposure to air or oxygen promotes β-Ta growth [34], [32]. To confirm this, we deposited tantalum films on oxidized and non-oxidized underlayers (see Appendix A, Table 3 for details).

Table 1. Tantalum film phase as a function of the substrate material Debye temperature (native – Ta deposited on an oxidized underlayer; in situ – Ta deposited immediately after the underlayer deposition).

| **Substrate** | **≈ $\theta_D$, K** | **Ta phase** | **ref. $\theta_D$** | **ref. Ta-phase** |
|---|---|---|---|---|
| Crystal Si | 650 | β | [69], [70], [71] | this work |
| Crystal $Al_2O_3$ | 1000 | β | [72], [73] | this work |
| Amorphous $SiO_2$ | 500 | β | [69], [74], [75] | this work |
| $NbO_X$ (native) | 640 | β | [62], [76] | this work |
| $CuO_X$ (native) | 610 | β | [63], [77] | [32] |
| $NiO_X$ (native) | 550 | β | [64], [78] | [32] |
| Nb (in situ) | 275 | α | [79], [80], [71] | this work |
| Cu (in situ) | 343 | unknown | [80], [71], [81] | - |
| Ni (in situ) | 450 | unknown | [80], [71], [81] | - |
| Mo (in situ) | 450 | α | [79], [80], [71] | this work |
| Al (in situ) | 428 | α | [80], [71], [81] | this work |
| Pt (in situ) | 237 | α | [79], [80], [71] | [34], [32] |
| Au (in situ) | 162 | α | [80], [71], [81] | [34], [32] |

In the case of heated silicon substrates, the β-Ta sublayer formation may be attributed to Si-Ta interdiffusion [65], [66], [67], which stabilizes the β-phase through mechanisms analogous to impurity stabilization in metallic glasses [68]. Moreover, the β-Ta sublayer does not lead to a large decrease in the $T_C$ of the films. As shown in Appendix D, even with a total film thickness of 20 nm, the $T_C$ is 3.77 K, while for a 150-nm-thick film it becomes 4.39 K.

### C. High-Q α-Ta/Si resonator measurements

In order to estimate quantitively the quality of a 150-nm-thick α-Ta films with 10-nm-thick β-phase sublayer, we fabricated superconducting coplanar quarter-wavelength resonators and measured their internal quality factors at single-photon regime (Figure 3, c). To reduce losses in the silicon substrate and at interfaces, which contribute to the overall resonator losses, we performed deep silicon etching (trench treating) in the resonator gaps [93], [94] and wet chemical post-processing to remove surface oxides at the metal-air (MA) and substrate-air (SA) interfaces. We also tested a combination of both methods, leaving the substrate-metal (SM) interface as the dominant one. In order to compare the proposed Ta technology to standard Al, we fabricated identical trenched resonators using polycrystalline and epitaxial SCULL aluminum (Single-crystalline Continuous Ultra-smooth Low-loss Low-cost) films [48]. All the coplanar resonators are designed for 50 Ohm impedance and compact footprint, where interfaces significantly contribute to the total losses. The dimensions were 4/7/4 µm (gap/line/gap) for planar resonators without substrate etching (Figure 3, d) and 4/10.5/4 µm for a 2 µm deep trenching ones (Figure 3, e). The designed frequency range is 4.5-5.1 GHz; 4 resonators are coupled to a common feedline with an external quality factor of $Q_C$ = (300 ± 100) ×$10^3$.

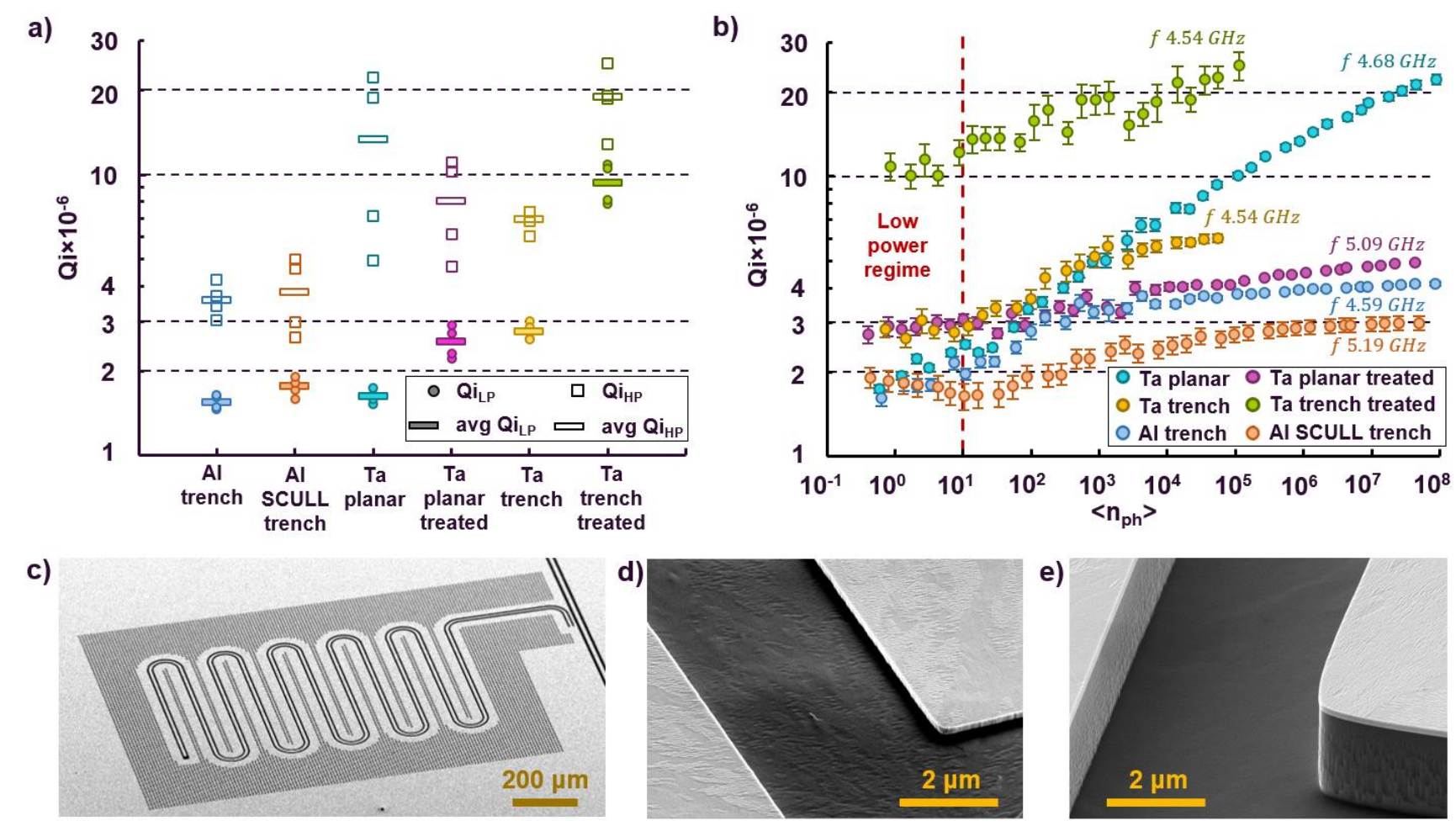

Figure 3. Ta and Al coplanar resonators investigation. (a) Internal quality factor at single-photon (filled circles) and high-power (empty squares) excitations: trench – resonators with substrate etching in the gap; treated – resonators with BOE treatment just before cryomeasurements; Al

SCULL – Single-crystalline Continuous Ultra-smooth Low-loss Low-cost aluminum film. Bars indicate the average $Qi_{LP}$ value for each group of resonators. (b) Internal quality factor at single-photon excitation for the best resonators from each group versus average photon number in the resonator. Error-bars indicate the range of Qi values obtained during 1 hours cryomeasurements. (c) SEM image of a trenched α-Ta resonator. (d) SEM image a planar α-Ta resonator gap. (e) SEM image a trenched α-Ta resonator gap.

All the resonators were fabricated on 5×10 mm² chips using α-Ta and Al films patterned by reactive-ion etching with $CF_4$ and $Cl_2$, respectively. After dicing, the chips were mounted in copper sample holders, and are measured in a dilution refrigerator below 10 mK. We implemented both infrared and magnetic shielding to minimize quasiparticle generation [82] and magnetic vortices. Then we measured the transmission coefficient $S_{21}$ using a vector network analyzer (VNA) and extracting internal quality factor $Q_i$ according to the circle-fit method described in Ref. [83]. The input and output lines with a total of 90 dB attenuation distributed across the dilution refrigerator stages were equipped with powder Eccosorb infrared filters [84], and low-pass filters. The output signals were amplified by impedance-matched parametric amplifier [85], [108] at 10 mK stage, following by HEMT amplifiers at 4 K stage and 300 K. Experimental setup and Qi measurement are described in Appendix C. By varying the drive signal power, we probed the resonator response across a wide range of photon population $\langle n_p \rangle$ from single-photon levels up to $10^7$ photons (Figure 3, b). At single-photon power, we observed quality factor $Qi_{LP}$ fluctuations exceeding 24% over several hours, therefore all reported $Qi_{LP}$ values represent time-averaged (over 1 hours) measurements rather than instantaneous.

First, we experimentally compared average $Qi_{LP}$ values for similar design trenched coplanar resonators fabricated using polycrystalline Al (~$1.6\times10^6$), SCULL epitaxial Al (~$1.8\times10^6$) and α-Ta ($2.7\times10^6$). One can notice, that epitaxial Al provides systematically higher $Qi_{LP}$ values, but this improvement looks neglible. While, α-Ta trenched resonators demonstrate almost 50% better performance in $Qi_{LP}$. This indicates lower losses in Ta oxides compared to Al oxides [10]. Further chemical treatment in Piranha and HF solutions allows increasing $Qi_{LP}$ up to ~$3\times10^6$ for planar and exceeding $10\times10^6$ for trenched α-Ta coplanar resonators, surpassing previously reported values [36], [39], [40], [41], [42], [37]. These results indicate, that the β-phase sublayer in 150-nm-thick α-Ta films does not introduce significant loss. The dominate source of loss for α-Ta films become the metal-air interface, which was substantially improved through post-processing treatments. The measured internal quality factors ($Qi_{LP}$ and $Qi_{HP}$) of compact coplanar resonators (Figure 3, a) confirm high α-phase purity of Ta films, demonstrating the great potential of α-Ta/Si for superconducting qubits and quantum applications. In Appendix E, the loss channels in the resonators were analyzed separately.

We also note that trench Ta-resonators enter a nonlinear regime when applied with a power greater than $10^5$ photons. This effect has been observed by other researchers [95], [96], [97] and can be related to experimental setup, external fields, the motion of magnetic vortices in the superconductor [98], etc. Our resonators are provided with magnetic and infrared shielding, but this proved insufficient. Furthermore, judging by the RRR value of our films (≈ 10) and the (110) crystal orientation, our films are in the dirty limit of superconductivity, which should lead to vortex pinning [98]. Nevertheless, nonlinearity persists at high applied powers. Furthermore, the $Qi_{LP}$ decreases immediately with increasing temperature, while our planar resonators exhibit an increase in $Qi_{LP}$ up to 500 mK (Appendix E). We suggest that the nonlinear behavior of tantalum resonators is not related to the resonator type (we observed nonlinearity for planar resonators too), but occurs randomly, even for resonators obtained from the same wafer, which has also been observed by other researchers [40], [95]. Thus, this effect is observed for both planar and trench Ta-resonators; for both sapphire as a substrate and silicon, so the nonlinear behavior of tantalum resonators requires separate careful study.

## III. CONCLUSION

In summary, we have experimentally confirmed a phase selection mechanism for the system Si/Ta, which is determined by the rate of adatoms thermalization. Furthermore, we demonstrated that during Ta sputtering onto silicon substrates at elevated temperature, a sub-10-nm-thick sublayer of β-Ta is first formed, following by α-Ta nucleation at discrete sites and further thicker film growth. Our results demonstrate that well-known silicon substrate pretreatment (removing surface adsorbates) itself don't allow α-Ta formation without substrate heating. It confirms a crucial role of a thermal activation for α-phase tantalum growth on both silicon and sapphire substrates. We identified critical formation temperatures of 450°C for silicon and 400°C for sapphire. We attribute a key role in the phase selection mechanism in tantalum thin films to the Debye temperature of the substrate material. The higher the Debye temperature, the faster the adatom thermalization occurs, and the fast thermalization rate ensures the growth of the β-phase, which is consistent with the molecular modeling results in the literature.

For silicon substrates, we established that high α-phase purity with minimal roughness can be achieved across a wide argon pressure range (0.3-9.0 mTorr), provided the substrate temperature remains below 600°C. The resulting α-Ta films on silicon develop a distinctive microstructure featuring large "flower-like" grains averaging 30-40 μm in size. These grains exhibit a preferred (110) orientation along the vertical axis while remaining randomly oriented in-plane, as confirmed by EBSD and XRD analysis. The film surface presents an intriguing combination of nanoridge structures (Sq ≈ 16.9 Å) and smooth regions (Sq ≈ 4.2 Å), whose formation mechanism remains unclear. Our investigation of the growth process at 500°C revealed a two-stage growth: initial deposition of a 10 nm β-Ta sublayer (ρ ≈ 150 μΩ×cm) followed by α-Ta nucleation (ρ ≈ 17 μΩ×cm) at discrete sites. By 20 nm thickness, the α-phase completely dominates the film without β-Ta recrystallization, a phenomenon we attribute to Ta-Si interdiffusion stabilizing the β-phase and influencing the eventual flower-grain morphology. It results in ranging the critical temperature of α-Ta films from 3.77 K to 4.39 K for the total thickness from 20 to 150 nm, respectively.

However, we claim that β-Ta sublayer has no dramatic effect on the superconducting films performance, demonstrating trenched α-Ta/ β-Ta /Si coplanar resonators with the state-of-the-art internal quality factor at single-photon power. We fabricated both planar and 2-μm-deep trenched λ/4 tantalum coplanar resonators with remarkable internal quality factors in the single-photon regime of $3\times10^6$ for and exceeding $10\times10^6$, correspondingly. Notably, that trenched Ta resonators outperformed their polycrystalline and epitaxial aluminum counterparts by a factor of five, while the β-Ta sublayer showed no measurable effect on resonator internal quality factor at single-photon excitations.

The samples are fabricated at the BMSTU Nanofabrication Facility (Functional Micro/Nanosystems, FMNS REC, ID 74300).

## APPENDIX A: The investigation of the substrate pretreatment influence on the tantalum films phase composition

It is well known that α-Ta typically forms when substrates are heated above 400°C [30], [31], [32]. While this heating effect is often attributed to surface cleaning through adsorbate desorption, our study systematically investigated how different substrate pretreatments affect the phase composition of 150-nm Ta films, as determined by resistivity measurements and EBSD analysis. We further verified these findings through XRD characterization of films deposited at both 20°C and 500°C (Figure 4). Experimental results are summarized in Table 2.

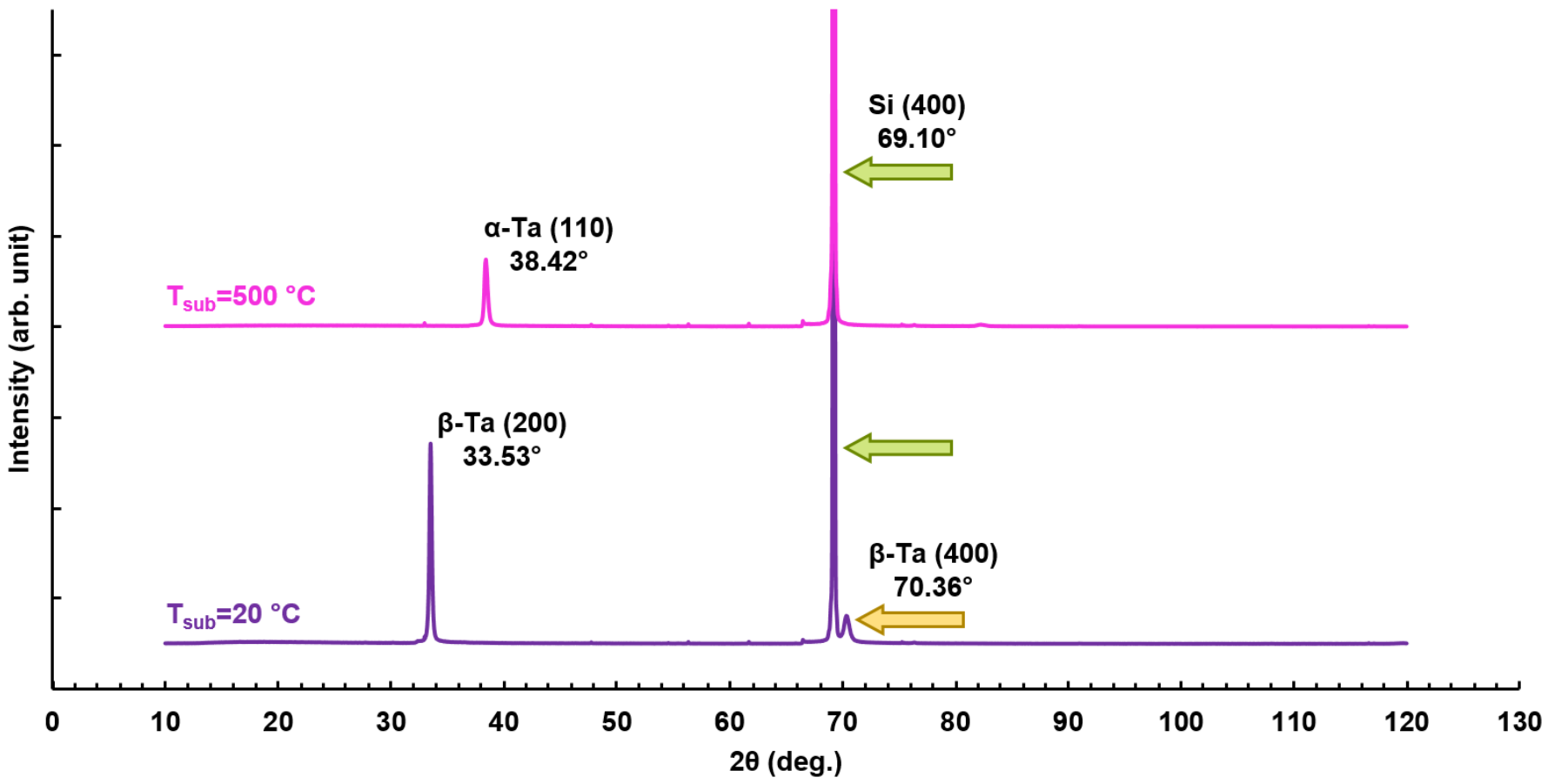

Figure 4. XRD spectra of Ta-films deposited at 20 and 500 °C

We examined two deposition conditions: unheated substrates (Group "C" in the Table 2) and heated substrates (Group "H" in the Table 2). Various surface preparation methods were evaluated, including: HF dipping for native oxide removal (HF dip in the Table 2), thermal degassing at 300°C for 3 hours (Degas. in the Table 2), high-temperature treatment at 500°C for 30 minutes and further cooling (500 + cooling in the Table 2), argon plasma cleaning for 15 minutes prior to deposition (Descum in the Table 2).

Table 2. Silicon substrate pretreatment and corresponding resistivity (μΩ×cm) and phase of tantalum thin films according to EBSD data. Sample "C" is unheated (cold) substrates; Sample "H" is heated substrate; "HF dip" is HF dipping for native oxide removal; "Degas." is thermal degassing at 300°C for 3 hours; "Descum" is argon plasma cleaning for 15 minutes prior to deposition; "$T_{heat}$, °C" is temperature of heating during deposition; "500+cooling" means high-temperature treatment at 500°C for 30 minutes and further cooling before deposition.

| Sample | HF dip | Degas. | Descum | $T_{heat}$, °C | ρ, μΩ×cm | Phase |
|---|---|---|---|---|---|---|

| C1 | - | + | - | - | 162.0 | β |
|---|---|---|---|---|---|---|
| C2 | + | - | - | - | 158.4 | β |
| C3 | + | + | - | - | 160.8 | β |
| C3 | + | + | + | - | 154.8 | β |
| C4 | + | - | + | - | 160.2 | β |
| C5 | + | - | - | 500+cooling | 158.5 | β |
| H1 | + | - | - | 400 | 162.7 | β |
| **H2** | **+** | **-** | **-** | **450** | **16.8** | **α** |
| **H3** | **+** | **-** | **-** | **500** | **17.1** | **α** |

All pretreatment methods produced films with resistivities exceeding 150 μΩ×cm, corresponding to pure β-Ta as confirmed by EBSD. Importantly, only deposition above critical temperatures yielded pure α-Ta with resistivities below 18 μΩ×cm. For silicon substrates, thermal activation remained essential for α-Ta formation regardless of surface treatment. For silicon substrates, thermal activation proves critical for α-Ta formation, with distinct threshold temperatures of 450°C for silicon and 400°C for sapphire. Figure 5, a illustrates the characteristic dependence of Ta film resistivity on substrate temperature. All depositions were performed using consistent magnetron sputtering parameters as specified in the main text.

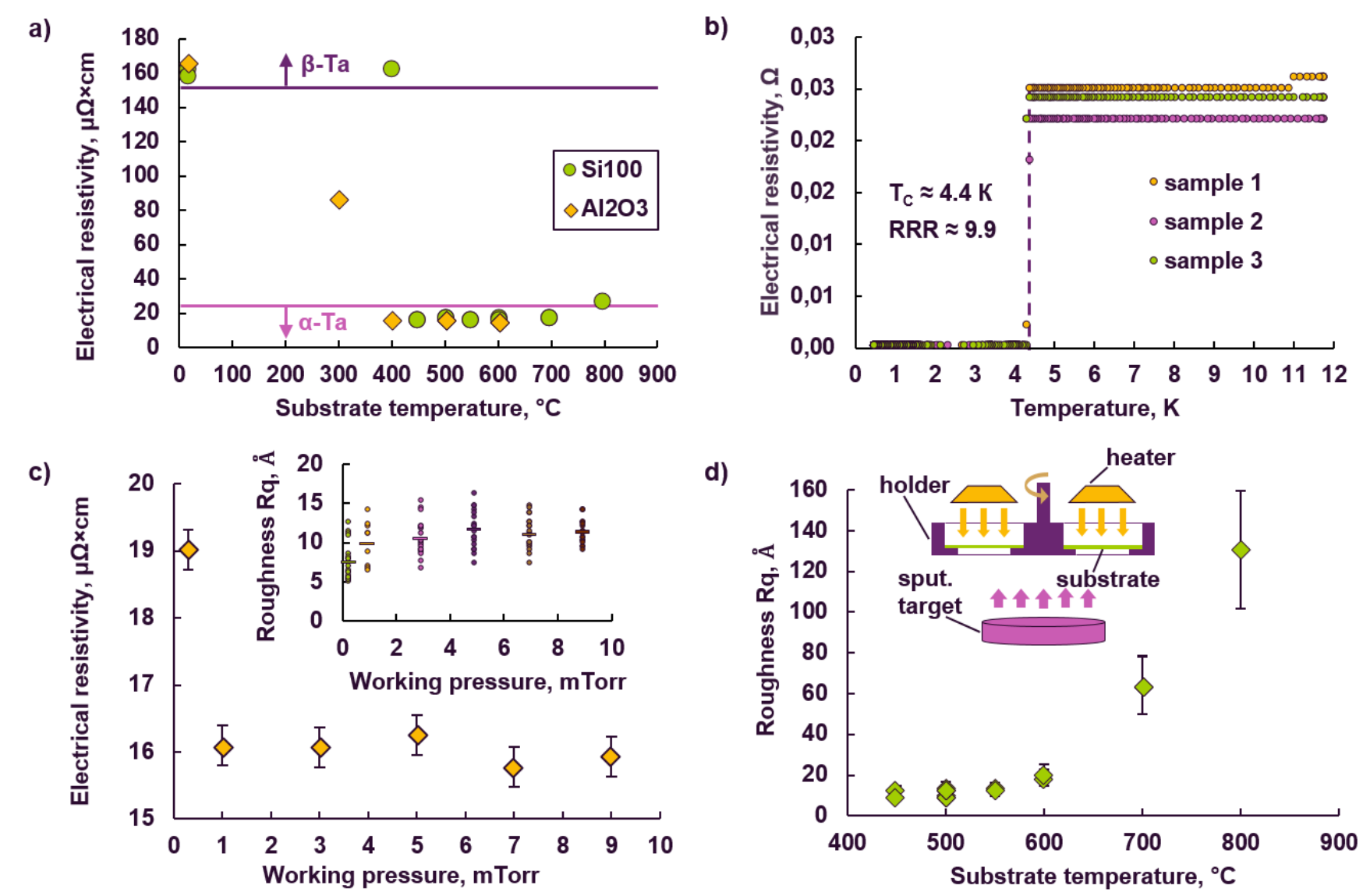

Figure 5. Study of Ta deposition regimes: (a) Resistivity of tantalum films versus substrate temperature during deposition. (b) Resistivity of Ta films on Si versus temperature measured in a dilution refrigerator. (c) Resistivity of Ta films on Si heated to 500 °C versus working pressure $p_{Ar}$; inset shows the dependence of roughness Rq on $p_{Ar}$. (d) Ta-films on Si versus substrate temperature during deposition; inset shows the magnetron sputtering deposition scheme.

Table 3 shows the resistivity values of tantalum films deposited on Mo, Nb and Al underlayers using argon plasma cleaning and heating to 500 °C.

Table 3. Resistivity (μΩ×cm) of tantalum films deposited on 50 nm Nb, Mo and Al underlayers and without an underlayer, under different surface conditions: **in situ** – Ta deposited immediately after underlayer deposition; **native** – Ta deposited on air-oxidized underlayer; **descum** – Ta deposited on an air-oxidized underlayer cleaned in argon plasma for 15 minutes just before deposition; **heat** – Ta deposited on an air-oxidized underlayer heated to 500 °C just before deposition. Resistivity values corresponding to α-Ta are indicated in green, and those corresponding to β-Ta are indicated in orange.

| surface / substrate | in situ | native | descum | heat |
|---|---|---|---|---|
| **Si** | - | 156.0 | 154.3 | 15.3 |
| **Si/Nb** | 20.4 | 139.8 | 20.2 | 17.3 |
| **Si/Mo** | 20.3 | 19.5 | 20.3 | 16.8 |
| **Si/Al** | 20.2 | 145.6 | 22.4 | - |

It can be seen that in the case of Al and Nb, deposition on a freshly deposited underlayer results in the growth of the α-Ta, but when the underlayer is air-oxidized, the β-Ta is formed. However, if the air-oxidized underlayer is etched (around 15 nm) using argon plasma cleaning, the α-Ta is formed again. At the same time, only heating allows the α-Ta to be deposited on a silicon substrate, which confirms that heating is the energy condition for the formation of the alpha-phase of tantalum. Additionally, the growth of α-Ta on the Nb underlayer oxidized in air and heated confirms the hypothesis about the influence of the Debye temperature of the substrate material on phase selection in Ta films, which is presented in the main text.

## APPENDIX B: The investigation of the magnetron sputtering parameters influence on the α-Ta/Si phase composition and roughness

For some α-Ta applications, such as quantum circuits, not only the phase composition of Ta films is important but also their roughness, which should be minimized [86]. There is evidence of a significant effect of process pressure on the phase composition of Ta films formed on unheated substrates [44], [45]. We evaluated the effect of the argon working pressure $p_{Ar}$ in the range from 0.3 to 9.0 mTorr on the Ta-films resistivity $\rho_{Ta}$ and roughness Rq of the Ta films. The dependence of $\rho_{Ta}$ on $p_{Ar}$ is shown in Figure 5, c: $\rho_{Ta}$ shows minimal dependence on $p_{Ar}$ ranging from 16.0 to 19.1 μΩ×cm, indicating no influence on the phase of the films. The dependence of Rq on $p_{Ar}$ is shown in the inset of Figure 5, c: across the full range, the Rq varies from 5.0 to 15.6 Å with a minimum average value of about 7.0 Å at $p_{Ar}$ of 0.3 mTorr.

We found that the substrate temperature $T_{heat}$ during α-Ta film formation has a significant effect on the roughness, as shown in Figure 5, d. In the $T_{heat}$ range from 450 to 600 °C, Rq does not exceed 20.0 Å. However, with further increases in $T_{heat}$, Rq rises significantly, reaching an average value of 132.2 Å at a substrate temperature of 800 °C. The resistivity $\rho_{Ta}$ across the entire $T_{heat}$ range does not exceed 18.0 μΩ×cm. However, for $T_{heat}$ of 800 °C $\rho_{Ta}$ exceeds 27.0 μΩ×cm, which, combined with the significant increase in Rq, indicates the poor quality of the α-Ta films formed at 800 °C.

Figure 5, b shows the electrical resistivity measurements during cooling from room temperature down to 10 mK in the dilution refrigerator for three α-Ta films fabricated using slightly different deposition regimes. The critical temperature for all samples is approximately 4.4 K, and the residual resistance ratio (RRR) is about 9.9. Compared with literature data [38], [39], [41], [46], [42] this suggests a pure high-Q tantalum α-phase.

It has been reported that tantalum films can undergo a β→α phase transition when they reach a large thickness during deposition without substrate heating [87], [88], [89]. However, we found no significant difference in the resistivity $\rho_{Ta}$ of films with thicknesses of 150, 300, and 1200-nm.

We would also like to note that for β-Ta films on silicon, we did not detect the superconducting transition, the film remained in the normal state down to 10 mK, with an RRR value of about 0.95. In terms of magnetron sputtering parameters, we found that process pressures in the range of 0.3–9.0 mTorr had no significant effect on film resistivity or roughness. The substrate heating temperature, however, proved to be a more important factor. We observed a significant increase in resistivity for $T_{heat}$ 800 °C and a dramatic increase in roughness Rq for $T_{heat}$ values of 700 and 800 °C. Therefore, in the case of silicon substrates, we recommend selecting a substrate temperature in the range of 450 °C to 600 °C.

### APPENDIX C: Measurement setup

To minimize resonator losses induced by non-equilibrium quasiparticles and magnetic vortex displacement, we mount the samples inside several layers of shielding (Fig. 6). Specifically, we anchor the PCB-mounted sample directly to a copper cold finger connected to the mixing chamber of the dilution refrigerator. The sample is then enclosed in a copper can, the inner surface of which is coated in a mixture of Stycast 2850 FT and silicon carbide granules with diameter 1000 µm. This can be enclosed in a second aluminum can. This is finally enclosed in one layer of cryogenic magnetic shielding (1-mm-thick Cryoperm). Coaxial cables entering the sample holder were pasted into the lid of the inner layer of IR shielding to reduce the impact of open holes on the shielding effectiveness. Extra radiation shielding is provided by in-house inline Eccosorb infrared filters in the input coaxial line mounted outside the magnetic shields at the mixing chamber stage.

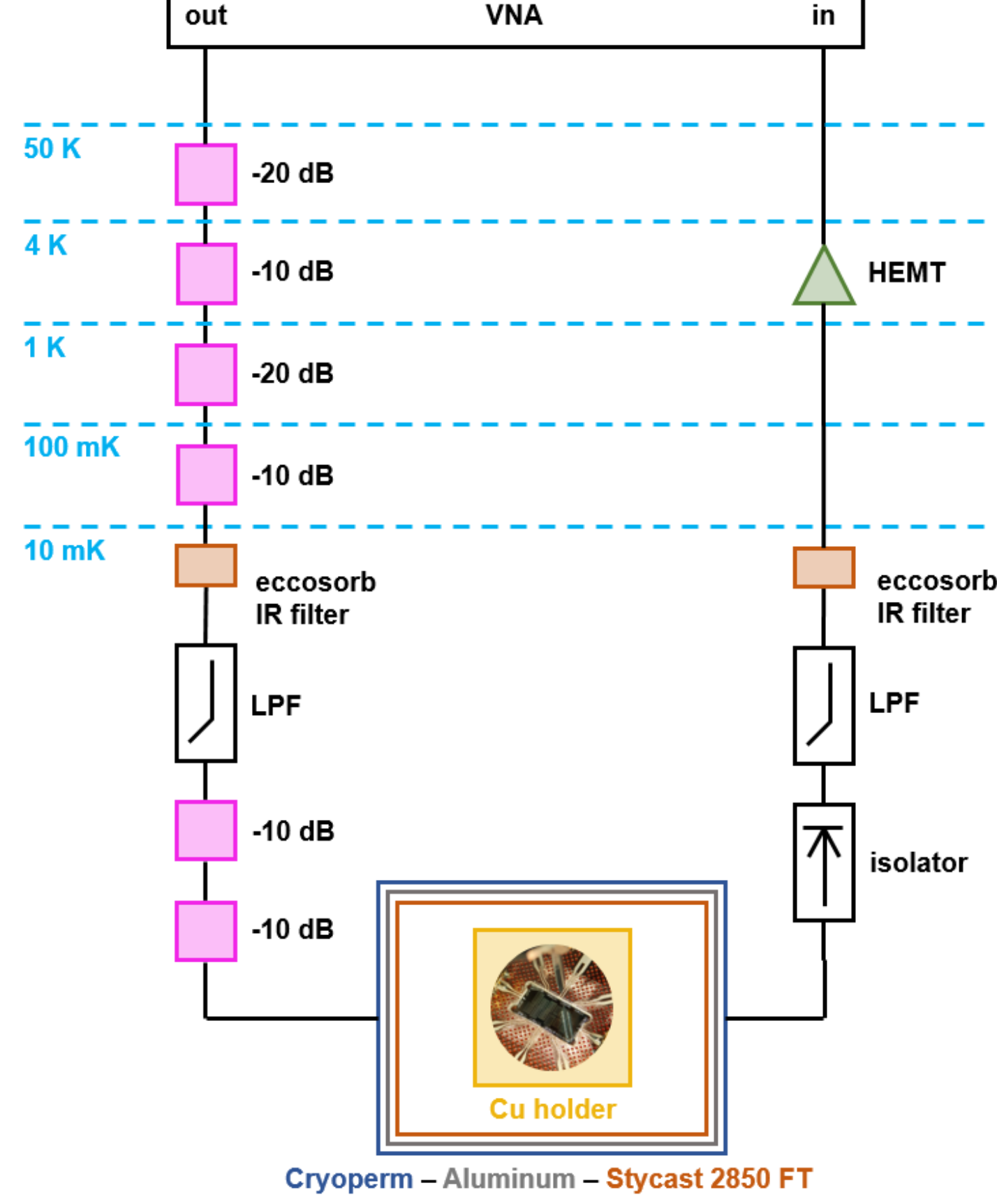

Figure 6. Wiring diagram of a measurement setup for the samples

All measurements are performed by means of a vector network analyzer (VNA) Rohde&Schwarz ZNB20 in the hanger mode [90]. The measurement procedure performs $S_{21}$ magnitude response measurement in the frequency range of 200-500 kHz around the resonance for each resonator. The resonance peak is located in the center of chosen range. We perform sweep routine of magnitude response from low-powers to the high power and extract the resonant frequency $f_{res}$, loaded quality factor $Q_l$, coupling quality factor $Q_c$ and internal quality factor $Q_i$ using opensource package described in detail in [83]. The power limits are chosen experimentally, so that the low-power corresponds to the single-photon regime, and the high-power does not put the resonator into nonlinear regime.

### APPENDIX D: Effect of film thickness on $T_C$

Tantalum films with thicknesses of 20, 30, 60 and 100 nm were additionally formed at a substrate temperature of 500 °C for evaluating the effect of film thickness on the critical temperature $T_C$ and, consequently, on the kinetic inductance. The results of $T_C$ measurements are shown in Figure 7. The vertical axis shows the normalized resistance (R at 7.5 K). The normalized resistance value for 30, 60, 100 and 150 nm thickness was multiplied by 0.9, 0.8, 0.7 and 0.6, respectively, for visibility. The $T_C$ was defined as the middle of the region between normal conductivity and superconductivity. The $T_C$ for films of 20, 30, 60, 100 and 150 nm was 3.77, 4.06, 4.27, 4.31 and 4.39 K, respectively.

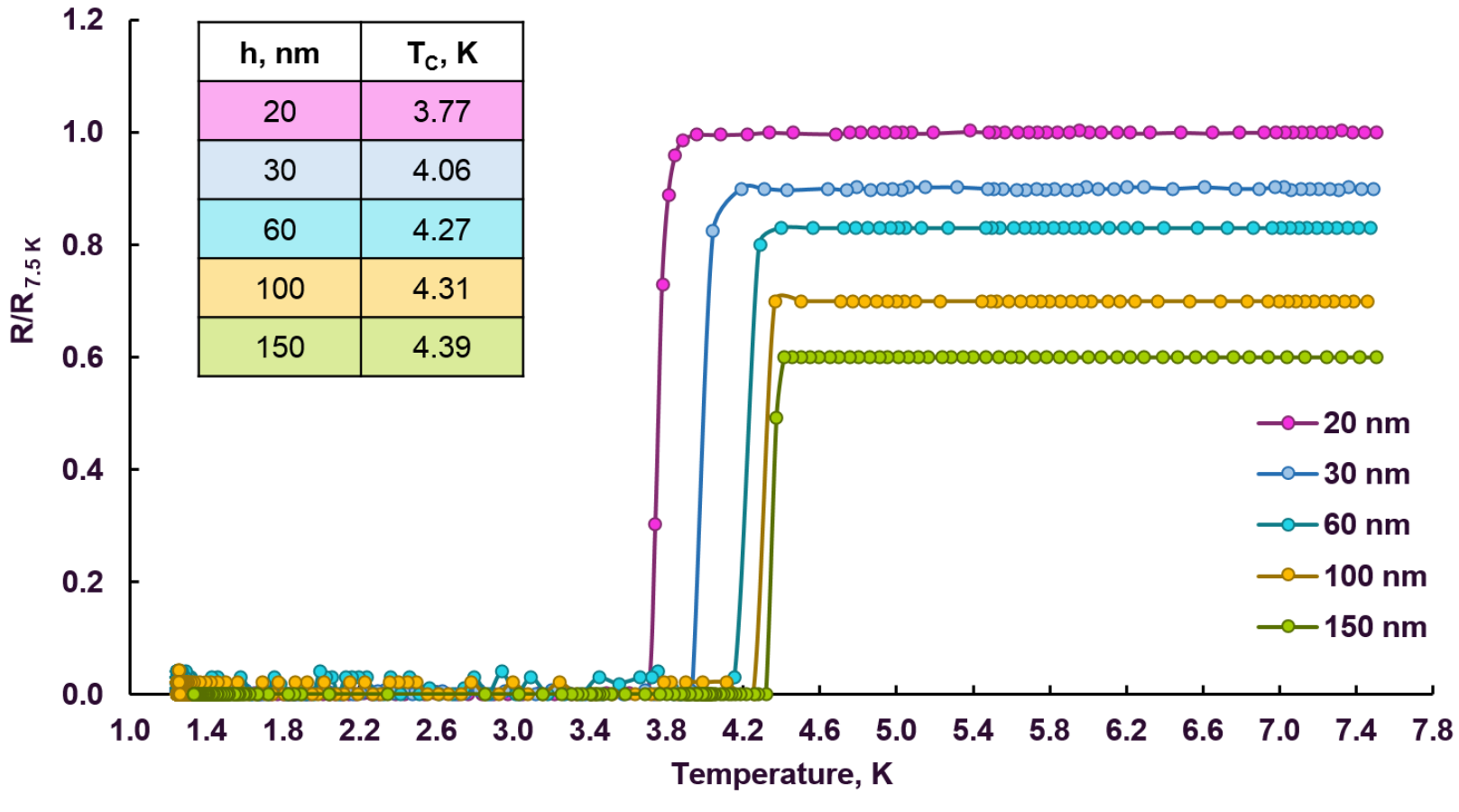

| h, nm | $T_C$, K |
|---|---|
| 20 | 3.77 |
| 30 | 4.06 |
| 60 | 4.27 |
| 100 | 4.31 |
| 150 | 4.39 |

Figure 7. Normalized resistance of Ta-films with different thickness. The normalized resistance value for 30, 60, 100 and 150 nm thickness was multiplied by 0.9, 0.8, 0.7 and 0.6, respectively, for visibility

In the work [91], tantalum films with a thickness of 40, 80 and 100 nm, formed on a niobium underlayer, were studied, and the $T_C$ for them was 3.9, 4.2 and 4.5 K, respectively. Similarly, in [99], α-Ta films were deposited on c-plane sapphire at 650°C, and for films with thicknesses of 19.3, 32.0, and 76.0 nm, the $T_C$ was 3.0, 3.5, and 3.9 K, respectively. In this case, the Ta-O interface layer was only 0.8 nm. Clearly, the reduction in $T_C$ for our α-Ta/Si films is even smaller than that for films of similar thickness deposited on a Nb underlayer and on a sapphire substrate. Consequently, the presence of a 10-nm β-Ta sublayer does not lead to an additional reduction in $T_C$, i.e., it does not increase the film's kinetic inductance. The observed pattern of decrease in $T_C$ is reminiscent of that for Nb films [100], [101], and one can expect the influence of quantum-size effects on $T_C$ of Ta films with a further decrease in thickness to less than 20 nm [100]. Thus, the developed technology is a reliable method for forming tantalum films for superconducting quantum circuits in a wide range of film thicknesses, including for quantum memory devices [109], [110].

### APPENDIX E: Resonator measurements at variable T

To disentangle the sources of losses in superconducting resonators, we conducted additional measurements of the resonators over a range of temperatures. The dependence of losses on power and temperature is described by three loss mechanisms: losses caused by TLS ($Q_{TLS}$), equilibrium quasiparticles ($Q_{QP}$) and separate loss channels independent on power and temperature ($Q_{other}$). The TLS losses are parameterized by the following function [92]:

$$Q_{TLS}(n_{ph},T) = Q_{TLS,0}\frac{\sqrt{1+\left(\frac{{n_{ph}}^{\beta_2}}{DT^{\beta_1}}\right)\tanh\left(\frac{\hbar\omega}{2k_BT}\right)}}{\tanh\left(\frac{\hbar\omega}{2k_BT}\right)}, \tag{E1}$$

while the losses caused by quasiparticles are:

$$Q_{QP}(T) = Q_{QP,0}\frac{e^{\Delta_0/k_BT}}{\sinh\left(\frac{\hbar\omega}{2k_BT}\right)K_0\left(\frac{\hbar\omega}{2k_BT}\right)}, \tag{E2}$$

where $\omega$ is the angular resonant frequency of the resonator; $n_{ph}$ is the number of photons in the resonator cavity; $Q_{TLS,0}$ is the inverse linear damping coefficient due to TLS; $D, \beta_1, \beta_2$ are parameters characterizing TLS saturation; $Q_{QP,0}$ is the inverse linear damping coefficient due to quasiparticles; $\Delta_0$ is the superconducting gap, calculated as $\Delta_0 = 1.764k_BT_c$; $T_c$ is the critical temperature of the film; $K_0$ is the zeroth order modified Bessel function of the second kind; $k_B$ is the Boltzmann constant; $\hbar$ is the reduced Planck constant. Thus, the complete model of the resonator's internal losses is described by the following formula:

$$\frac{1}{Q_{int}} = \frac{1}{Q_{TLS}(n_{ph},T)} + \frac{1}{Q_{QP}(T)} + \frac{1}{Q_{other}} \tag{E3}$$

Additionally, $Q_{TLS,0}$ and $Q_{QP,0}$ can be determined by measuring the shift in the resonator's central frequency as a function of temperature. The contribution to the frequency shift from TLS is described by the following model [92]:

$$\left(\frac{\delta f(T)}{f_0}\right)_{TLS} = \frac{1}{\pi Q_{TLS,0}} \mathrm{Re}\left[\Psi\left(\frac{1}{2}+i\frac{\hbar\omega}{2k_BT}\right) - \ln\left(\frac{\hbar\omega}{2k_BT}\right)\right], \qquad \text{(E4)}$$

while the contribution from the quasiparticles is:

$$\left(\frac{\delta f(T)}{f_0}\right)_{QP} = -\frac{\alpha}{2}\left(1-\sin\left(\phi(T,\omega)\right)\sqrt{\frac{\sigma_1(T,\omega)^2+\sigma_2(T,\omega)^2}{\sigma_1(0,\omega)^2+\sigma_2(0,\omega)^2}}\right), \qquad \text{(E5)}$$

where $\Psi$ is the complex digamma function; $\sigma_1$ and $\sigma_2$ are the real and imaginary parts of the complex conductivity; $\phi$ is the phase between the real and imaginary parts of the complex conductivity; $\alpha$ is the kinetic inductance fraction. Thus, the complete model for the resonator's frequency shift as a function of temperature is given by:

$$\frac{\delta f(T)}{f_0} = \left(\frac{\delta f(T)}{f_0}\right)_{TLS} + \left(\frac{\delta f(T)}{f_0}\right)_{QP}. \qquad \text{(E6)}$$

We measured three treated planar resonators made from a 150 nm-thick Ta film and simultaneously fitted the model from Eq. (1) using six fitting parameters: $Q_{TLS,0}$, $D$, $\beta_1$, $\beta_2$, $Q_{QP,0}$, $Q_{other}$ (see Figure 8, a). Additionally, we fitted the model from Eq. (6) using $Q_{TLS,0}$ and $\alpha$ as fitting parameters (see Figure 8, b).

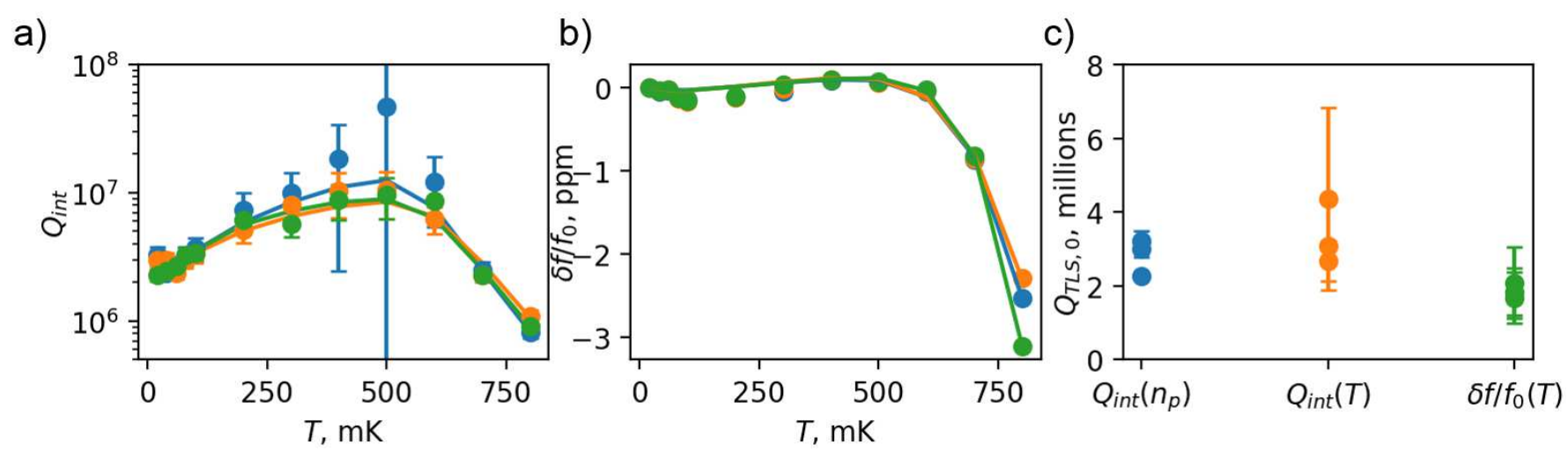

Figure 8. Results of resonator measurements at Variable Temperatures. The color coding represents different resonators. Dots correspond to experimental data, while solid lines indicate fitted models. (a) Internal quality factor as a function of temperature, measured at a single-photon power level. (b) Frequency shift versus temperature. (c) Comparison of different models for estimating TLS-induced losses.

$Q_{int}$ increases by more than a factor of 4 with temperature, peaking at approximately 8.5 million at 500 mK. At higher temperatures, $Q_{int}$ drops exponentially, consistent with the equilibrium quasiparticle model. The behavior of the $Q_{int}$ curve confirms that, in planar structures at low powers, $Q_{int}$ is TLS-limited, and suggests that the 10 million quality factors achieved in trenched treated resonators result from the removal of interfaces with high TLS losses. We summarize the fitted parameters in Table 4. Table 4. Fitted parameters for resonators measured at different temperatures.

| Model | Res.# | $Q_{TLS,0}$. | $Q_{QP,0}$, | $Q_{other}$ | $D$ | $\beta_1$ | $\beta_2$ | $\alpha$ |
|---|---|---|---|---|---|---|---|---|
| $\frac{1}{Q_{int}(T)}$ | 1 | 2.6e6 | 31.1e6 | 2.6e9 | 5.8 | 27.4 | 61.4 | - |
| | 2 | 3.8e6 | 43.7e6 | 20.9e6 | 27.4 | 0.9 | 13.2 | - |
| | 3 | 4.3e6 | 36.9 | 17.4e6 | 174.6 | 0.3 | 2.6 | - |
| $\frac{\delta f(T)}{f_0}$ | 1 | 3.0e6 | - | - | - | - | - | 1.6e-3 |
| | 2 | 2.8e6 | - | - | - | - | - | 0.8e-3 |
| | 3 | 3.7e6 | - | - | - | - | - | 0.8e-3 |

In Figure 8, c, we present the fitted $Q_{TLS,0}$ values obtained from the models in Eq. (3) and Eq. (6), as well as from the model describing $Q_{int}$ as a function of power [92]:

$$\frac{1}{Q_{int(n_{ph})}} = \frac{\tanh\left(\frac{\hbar\omega}{2k_B T}\right)}{Q_{TLS,0}\left(1+\frac{n_{ph}}{n_c}\right)^{\beta}} + \frac{1}{Q_{other}}, \quad \text{(E7)}$$

where $n_c$ is the critical photon number; $\beta$ is a fitting parameter. The results from all three models yield closely matching $Q_{TLS,0}$ values, confirming that TLS dominate the limiting factor in resonator quality factors.

## APPENDIX F: Morphology and resistivity of the β-Ta sublayer

We created a cross-section in a 100 nm thick α-Ta film using a focused ion beam (FIB) to confirm that the β-Ta sublayer persists in thicker films. Figure 9, a shows a cross-section image obtained in a FIB-system. The sublayer is clearly visible. We would like to emphasize that during cone-shaped growth of the α-Ta film, the base of the cone increases extremely rapidly as its height increases. At a film thickness of 10 nm, the base diameter is approximately 10 μm, and at a thickness of 15 nm, it is approximately 50 μm. In other words, the growth is cone-shaped, but it is unlikely that this geometric shape can be identified in cross-section.

To exclude island growth mechanism of α-Ta on the β-Ta sublayer and confirm cone-shaped growth, we performed an AFM study of a 15 nm thick tantalum film. The image was taken in such a way as to capture the growing α-Ta region and the surrounding β-Ta (Figure 9, b). The spatial resolution of the images is high enough to clearly observe the fine structure of the film (Figure 9, c). In addition, the boundary between the different zones of the sample is clearly visible. This boundary was identified as a step with a height of 0.87 Å. It is this boundary that creates the contour in the case of an optical image in a dark field (Figure 2, b). However, no additional spatial structure was observed. On the same film area surface potential distribution were studied by AFM in Kelvin Probe mode with frequency modulation on the second pass. To increase spatial resolution and sensitivity of the measurement non-resonant topography mode was used on the first pass. In this way topography images were taking simultaneously with surface potential distribution images. Using sharp non-coated conductive probe with different lift height for the second pass (from 10 to 200 nm) no image contrast was observed across the edge between dark and light zones.

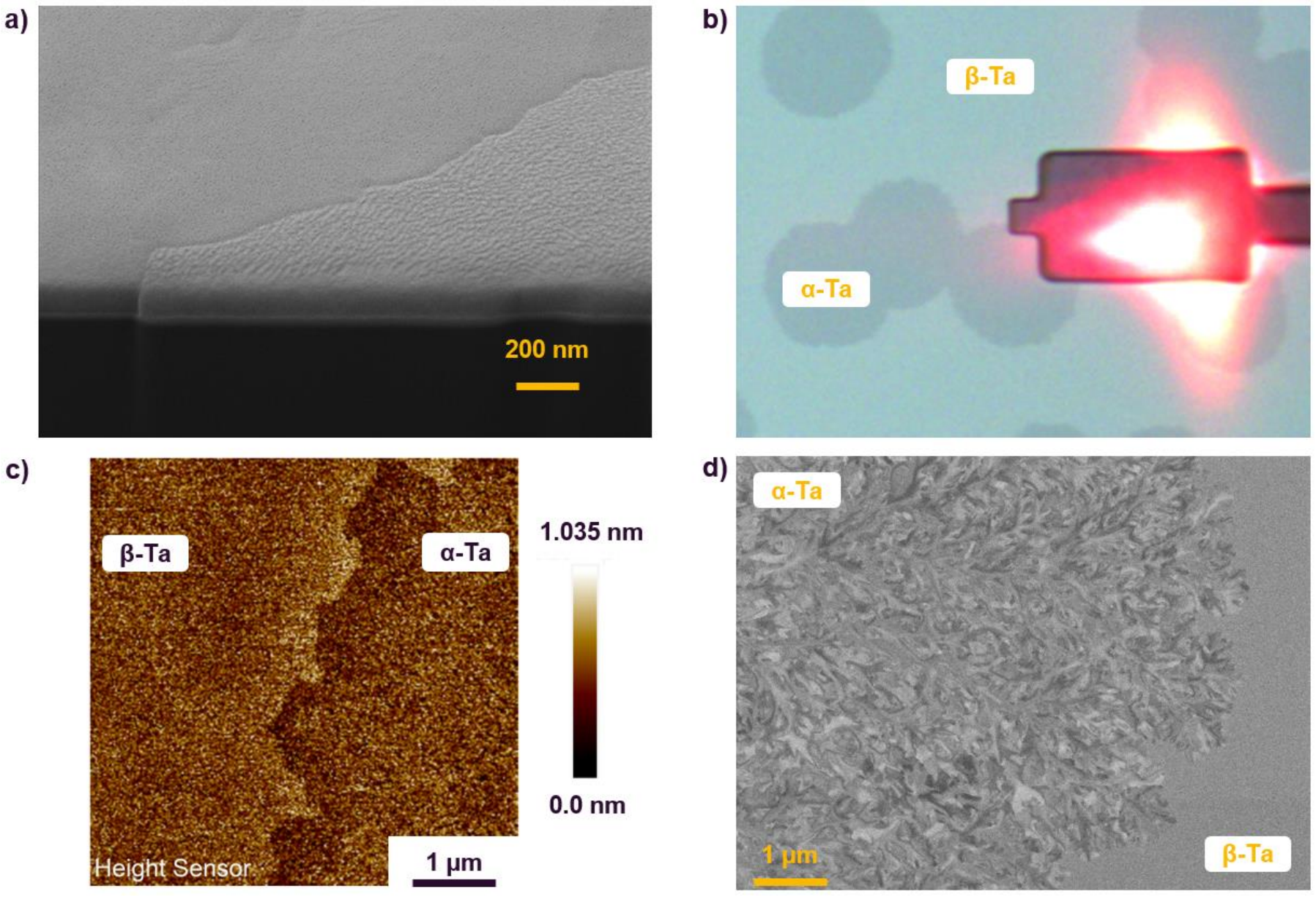

Figure 9. Investigation of the β-Ta sublayer morphology: (a) Cross-section image of 100 nm a-Ta film obtained using a FIB-system. (b) Optical images showing a AFM-tip position at the top end of the scan area. Dark (α-Ta) and bright (β-Ta) zones of the sample are clearly seen. (c) AFM-image of a 15 nm tantalum film at the boundary of the α-Ta and β-Ta regions. (d) SEM-image of a 15 nm tantalum film at the boundary of the α-Ta and β-Ta regions.

Remarkably, the structure of the α-Ta region in the SEM image (Figure 9, d) is more contrasting than the morphology of this region in the AFM image. We attribute this to the different electronic structures of the alpha and beta phases of tantalum, which provides differences in the optical properties and conditions for the escape of secondary electrons during SEM study. The morphology of the 15 nm Ta film doesn't have time to change significantly, so, with the exception of a barely visible boundary, we see no difference between the α-Ta and β-Ta regions. Clearly, there is no three-dimensional growth, and the alpha- and beta-tantalum regions grow upward simultaneously.

For our 5- and 10-nm-thick films, the resistivity is virtually identical to that of bulk β-Ta. However, it is known that the resistivity of metallic films typically increases up to threefold compared to the bulk material when going to ultrathin films due to a significant increase in the proportion of scattering at the surface and grain boundaries [102], [103]. These effects become more pronounced as the film thickness decreases below the electron mean free path [102]. We hypothesize, that the electron mean free path for β-Ta is significantly shorter than that of conventional metals, since the resistivity of β-Ta is an order of magnitude higher than that of conventional moderately conducting metals such as Nb [104], Mo [105], or α-Ta, and the residual resistance coefficient for β-Ta is approximately 0.95 (see Appendix B for details). Thus, we believe that the mechanisms that increase the resistivity of metals in ultrathin films have less influence on the resistance of β-Ta films. However, a sharp increase in the resistance of β-Ta with decreasing film thickness has been reported [106]. We suppose that this may be due to the unaccounted interfacial layer, which is why the actual β-Ta sublayer thickness in this study was smaller, resulting in a significant increase in the calculated resistivity. In order to properly address this point, we investigate the resistivity of β-Ta films with the thicknesses down to 5 nm. A series of β-Ta films with thicknesses of 5, 10, 15, 20, and 50 nm have been sputtered on silicon substrates without heating or an underlayer, and then are measured (Figure 10).

Figure 10. Experimental resistivity of β-Ta films on Si as a function of thickness.

The experimentally measured resistivity of these films was 159.7, 147.4, 146.2, 142.4, and 141.0 μΩ×cm, respectively. We've confirmed an increase in resistance from 141 to 159.7 μΩ×cm with decreasing the thickness for 50 to 5 nm. One can observe, that the resistance change is order of magnitude smaller than the difference between the resistances of the alpha and beta phases of tantalum. The slight increase in resistivity in our case may also be due to the presence of an unaccounted interfacial layer with a thickness of ≈ 0.5 nm, but no significant increase in resistivity was observed.

**Data availability**

The data that support the findings of this study are available within the article.

**Acknowledgements**

Technology was developed and samples were fabricated at the BMSTU Nanofabrication Facility (Functional Micro/Nanosystems, FMNS REC, ID 74300).

**Author contributions statement**

E.V.Z. proposed the growth mechanism guided by S.P.B. and I.A.R. E.V.Z., S.V.B., S.P.B., I.A.Ry., A.V.A. and I.A.R. planned, performed and analyzed the film growth experiments. E.V.Z. and A.R.M. designed of the coplanar resonators. E.V.Z., S.V.B., D.A.B., S.A.K., I.A.Ry. and I.A.R. developed of the technology and fabricate the coplanar resonators. A.R.M., A.I.I., E.I.M., V.I.P. and I.A.R. configured the cryogenic setup and control system; characterized the resonators. E.V.Z., E.A.K., S.V.B., D.A.F. and I.A.S. characterized tantalum films. E.V.Z., N.S.S., E.A.K. and I.A.R. wrote the manuscript with inputs from all co-authors. A.V.A. and I.A.R supervised the work.

**Additional information**

The authors declare no conflict of interest.